\documentstyle[prb,aps,amssymb,psfig,floats,amsmath]{revtex}

\begin{document}

\author{D. V. Dmitriev, V. Ya. Krivnov, and A. A. Ovchinnikov}
\address{Joint Institute of Chemical Physics of RAS, Kosygin str.4, 117977,
Moscow, Russia} 
\title{Exact ground states for a class of one-dimensional frustrated quantum
spin models }
\maketitle

\begin{abstract}
We have found the exact ground state for two frustrated quantum spin-1/2
models on a linear chain. The first model describes ferromagnet --
antiferromagnet transition point. The singlet state at this
point has double-spiral ordering.
The second model is equivalent to special case of the spin-1/2
ladder. It has non-degenerate singlet ground state with exponentially
decaying spin correlations and there is an energy gap. The exact ground
state wave function of these models is presented in a special recurrent form
and recurrence technics of expectation value calculations is developed.

\end{abstract}

\section{Introduction}

Last decade frustrated Heisenberg models have been a subject of intensive
studies$^{1-12}$. Of main interests are ground state properties with respect
to variations of exchange integrals and character of the phase transitions.
In particular, these properties may be important in the theory of high-$T_c$
superconductivity.

There is much interest in quantum spin systems with competing interactions
for which exact ground state can be constructed. The first example of such a
model has been given by Majumdar and Ghosh$^{13}$. They have considered $
s=\frac 12$ chain with antiferromagnetic nearest- and next-nearest neighbor
interactions and the strength of the second interaction is one half of the
first. The ground state of this model is two-fold degenerate, consists of
dimerized singlets and there is a gap in the spectrum of excited states.

Another example found by Affleck, Kennedy, Lieb and Tasaki$^{14}$ is an $S=1$
spin chain with special bilinear and biquadratic interactions (AKLT model).
This model has unique resonating-valence-bond ground state and ground state
correlations have exponential decay. Besides, there is the gap between
ground state and excited states. Further generalizations of the AKLT model
have been studied in a number recent papers$^{15}$.

In this paper we present two classes of $s=\frac 12$ chains for which exact
ground state wave function has a special recurrent form. These models have
competing ferro- and antiferromagnetic interactions and their ground states
can be either ferromagnetic or singlet depending on the relation between
exchange integrals. The first type of exactly solvable models is related to
the systems at the ferromagnet-antiferromagnet transition point when the
ferromagnetic and the singlet states are degenerate. The calculation of the
spin correlation function in the singlet ground state shows the spiral
magnetic order at this point.

The model of the second type has the nearest, next-nearest and
next-next-nearest neighbor interactions depending on one parameter. This
model has the non-degenerate singlet ground state for cyclic chains and for
a certain region of the parameter and its ground state properties are
similar to that of AKLT model. In other words, this model has all properties
of the ''Haldane scenario''$^{16}$, though it is the model with half-integer
spin. We note, however, that the considered model has two sites in an unit
cell and it is equivalent to the special case of a spin ladder. In some
limit this model reduces to the effective spin-1 chain for which the ground
state wave function coincides with that for the AKLT model.

The paper is organized as follows. In Section 2 we will consider the
model of the first type and describe the exact singlet ground state wave
function as well as details of the spin correlation function calculations.
Section 3 and Appendix present the study of the model of the second type.
The results of the paper will be summarized in Section 4.

\section{Frustrated spin chain at ferromagnet-antiferromagnet transition
points.}

\subsection{The exact ground state wave function.}

Let us consider $s=\frac 12$ spin model with nearest- and next-nearest
neighbor interactions given by the Hamiltonian
\begin{equation}
H=-\sum_{i=1}^M({\bf S}_{2i-1}\cdot{\bf S}
_{2i}-\frac 14)+J_{23}\sum_{i=1}^M({\bf S}_{2i}\cdot{\bf S}
_{2i+1}-\frac 14)+J_{13}\sum_{i=1}^N({\bf S}_i\cdot{\bf S}_{i+2}-\frac 14) ,
\label{1}
\end{equation}
with periodic boundary conditions and even $N=2M$.

If $J_{23}<1$, then the ground state of (1) is ferromagnetic (singlet) at $
\delta <0$ ($\delta >0$), where $\delta =J_{13}+\frac{J_{23}}{2(1-J_{23})}$
(Fig.1). The equation $\delta =0$ defines the line of transition points from
the ferromagnetic to the singlet state, when energies of these states are
zero. The model (1) along this line is given by the Hamiltonian depending on
the parameter $\nu $ ($\nu >0$):
\begin{figure}[t]
\unitlength1cm
\begin{picture}(11,8)
\centerline{\psfig{file=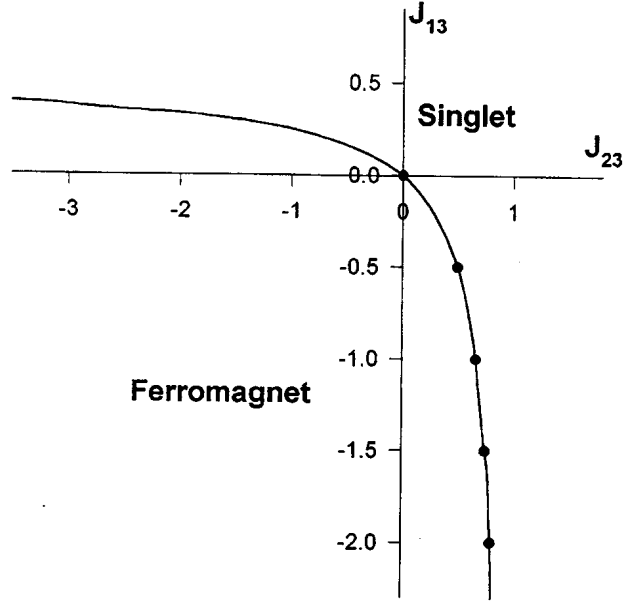,angle=0,width=9cm}}
\end{picture}
\caption[]{$T=0$ phase diagram of the  model (1). The solid line is the 
boundary between the ferromagnetic and singlet phases. Model (1) along this
line is given by Hamiltonian (2). Circles correspond to the special cases of
this Hamiltonian.}
\end{figure}
\begin{equation}
H=-\sum_{i=1}^M({\bf S}_{2i-1}\cdot{\bf S}_{2i}-\frac 14)
-(\nu -1)\sum_{i=1}^M({\bf S}_{2i}\cdot{\bf S}_{2i+1}-\frac 14)
+\frac{\nu -1}{2\nu }\sum_{i=1}^N({\bf S}_i\cdot{\bf S}_{i+2}-\frac 14) , \label{2}
\end{equation}
with periodic boundary conditions.

We note that the Hamiltonian (2) has a symmetry: its spectrum coincides with
the spectrum of $\widetilde{H}(\nu )$ obtained by the following
transformation
\[
\widetilde{H}(\nu )=(\nu -1)~H(\frac \nu {\nu -1}),\qquad \nu >1
\]

This transformation permutes the factors at the first and the second terms
in the Hamiltonian (2). Thus, due to the symmetry it is sufficient to
consider the range $0\leq \nu \leq 2$.

First, we will show that the ground state energy of (2) is zero. Let us
represent the Hamiltonian (2) as a sum of Hamiltonians $H_n$ of cells
containing three sites

\begin{equation}
H=\sum_{i=1}^M(H_{2i-1}+H_{2i}),  \label{3}
\end{equation}
where
\[
H_{2i-1}=-\frac 12({\bf S}_{2i-1}\cdot{\bf S}_{2i}-\frac 14)-\frac{\nu -1}2({\bf S
}_{2i}\cdot{\bf S}_{2i+1}-\frac 14)+\frac{\nu -1}{2\nu }({\bf S}_{2i-1}\cdot{\bf S}
_{2i+1}-\frac 14) ,
\]
\[
H_{2i}=-\frac 12({\bf S}_{2i+1}\cdot{\bf S}_{2i+2}-\frac 14)-\frac{\nu -1}2({\bf S
}_{2i}\cdot{\bf S}_{2i+1}-\frac 14)+\frac{\nu -1}{2\nu }({\bf S}_{2i}\cdot{\bf S}
_{2i+2}-\frac 14)
\]

Eigenvalues of each $H_n$ are
\[
\lambda _1=\lambda _2=0,\qquad \lambda _3=\frac{\nu ^2-\nu +1}{2\nu }>0
\]

We will present a singlet wave function which is the exact one of each $H_n$
with zero energy and, therefore, it is the exact ground state wave function
of (2). This function has a form
\begin{equation}
\Psi _0(M)=P_0\Psi _M,  \label{4}
\end{equation}
where
\begin{eqnarray}
\Psi _M &=&(s_1^{+}+\nu s_2^{+}+\nu s_3^{+}\ldots +\nu s_N^{+})(s_3^{+}+\nu
s_4^{+}\ldots +\nu s_N^{+})\ldots  \nonumber \\
&&\times (s_{2n-1}^{+}+\nu s_{2n}^{+}\ldots +\nu s_N^{+})\ldots (s_{N-1}^{+}+\nu
s_N^{+})\mid 1,2,\ldots N\rangle ,  \label{5}
\end{eqnarray}
where $s_i^{+}$ is the $s=\frac 12$ raising operator.

Eq.(5) contains $M=\frac N2$ operator multipliers and the vacuum state $\mid
1,2,\ldots N\rangle $ is the state with all spins pointing down. The function $
\Psi _M$ is the eigenfunction of $S_z$ with $S_z=0$ but it is not the
eigenfunction of ${\bf S}^2$. $P_0$ is a projector onto the singlet state.
This operator is$^{17}$
\[
P_0=\frac 1{8\pi ^2}\int_0^{2\pi }d\alpha \int_0^{2\pi }d\beta \int_0^\pi
\sin \gamma d\gamma ~e^{i\alpha \stackrel{\wedge }{S_z}}e^{i\gamma \stackrel{
\wedge }{S_x}}e^{i\beta \stackrel{\wedge }{S_z}},
\]
where $\stackrel{\wedge }{S}_{x(z)}$ are components of the total spin
operator.

The function $\Psi _M$ contains components with all possible values of spin $
S$ ($0\leq S\leq M$) and, in fact, a fraction of the singlet is
exponentially small at large $N$. This component is filtered out by the
operator $P_0$.

It is not difficult to check that
\begin{equation}
H_n\Psi _M=0  \label{6}
\end{equation}
for $n=1,\ldots (N-2)$ and, therefore, the ground state energies for all values
of spin $S$ of an open chain describing by the Hamiltonian
\begin{equation}
H_{\rm op}=\sum_{n=1}^{N-2}H_n  \label{7}
\end{equation}
are zero.

The operators $H_{N-1}$ and $H_N$ do not give zero acting on $\Psi _M$ but
\begin{equation}
H_{N-1(N)}P_0\Psi _M=P_0H_{N-1(N)}\Psi _M=0  \label{8}
\end{equation}

The latter equation can be easily checked using the fact the first bracket
in (5) can be replaced by $(1-\nu )s_1^{+}$ under the projector $P_0$.

Eqs.(6) and (8) mean that $\Psi _0(M)$ is the exact singlet ground state
wave function of (2) for any $M$ and the ground state energy is zero. We
note that the ground state energy coincides with its exact lower bound
because $H$ is the sum of non-negative defined operators (Eq.(3)). Of
course, the trivial ferromagnetic state has zero energy as well.

In particular case, $\nu =2$, when $J_{12}=J_{23}=-1$ and $J_{13}=\frac 14$,
another form of the exact singlet ground state wave function has been found
in Ref.18. It reads
\[
\Psi =\sum [i,j][k,l][m,n]\ldots ,
\]
where $[i,j]$ denotes the singlet pair and the summation is made for any
combination of spin sites under the condition that $i<j,k<l,m<n\ldots$.
However, it is not clear how this function can be generalized to $\nu\neq 2$.

The following general statements relevant to the Hamiltonian (2) can be
proved:

1). The ground states of open chains described by (7) in the sector with
fixed total spin $S$ are non-degenerate and their energies are zero.

2). For cyclic chains the ground state in the $S=0$ sector is non-degenerate
and has momentum $\pi $. The ground state energies for $0<S<M$ are non-zero.

3). The singlet ground state wave function for open and cyclic chains
coincide with each other.

4). The singlet ground state wave function $\Psi _0(M)$ is superstable$^{19}$
with respect to any cell operator $H_n$, i.e. $\Psi _0(M)$ is the ground
state wave function of the Hamiltonian $H+\lambda H_n$ for $-1<\lambda
<\infty $.

\subsection{The norm of the ground state wave function.}

Let us return to the problem of the projection of the function $\Psi _M.$ As
one can see from Eq.(5) the function $\Psi _M$ satisfies a recurrent equation
\begin{equation}
\Psi _M=\left[ (s_1^{+}+\nu s_2^{+})+\nu S^{+}(N-2)\right] \mid 1,2\rangle
\Psi _{M-1},  \label{9}
\end{equation}
where $\Psi _{M-1}$ is the function (5) for the system of $(N-2)$ spins on
sites $3,4,\ldots N$ and $S^{+}(N-l)=\sum_{i=l+1}^N\,s_i^{+}.$

In principle, it is possible to generate $\Psi _0(M)$ starting from $M=1$
and using Eqs.(4) and (9). However, it is more convenient to obtain the
recurrent formulae for expectation values (norm and correlators) with
respect to the function $\Psi _0(M)$ directly.

First, we consider a norm of $\Psi _0(M)$ which has a form
\begin{equation}
G_M=\langle \Psi _0(M)\Psi _0(M)\rangle =\langle \Psi _MP_0\Psi _M\rangle
=\frac 12\int_0^\pi \Phi _M(\gamma )\sin \gamma d\gamma ,  \label{10}
\end{equation}
where
\[
\Phi _M(\gamma )=\Phi _M(\gamma _1,\gamma _2)\mid _{\gamma _1=\gamma
_2=\gamma }
\]
and
\begin{equation}
\Phi _M(\gamma _1,\gamma _2)=\langle \Psi _M\exp (i\frac{\gamma _1}
2S^{+}(N)+i\frac{\gamma _2}2S^{-}(N))\Psi _M\rangle  \label{11}
\end{equation}

Commuting operators in Eq.(11) and using the fact, that $S_z\Psi _M=0,$ we
rewrite Eq.(11) in a form
\begin{equation}
\Phi _M(\gamma _1,\gamma _2)=\langle \Psi _M\exp (izS^{-}(N))\exp
(iz^{\prime }S^{+}(N))\Psi _M\rangle , \label{12}
\end{equation}
where
\[
z=\sqrt{\frac{\gamma _2}{\gamma _1}}\tan \frac{\sqrt{\gamma _1\gamma _2}}
2,\qquad z^{\prime }=\sqrt{\frac{\gamma _1}{\gamma _2}}\frac{\sin \sqrt{
\gamma _1\gamma _2}}2
\]

It follows from Eqs.(9) and (12) that the function $\Phi _M(\gamma _1,\gamma
_2)$ satisfies the equation
\begin{eqnarray*}
\Phi _M(\gamma _1,\gamma _2) &=&[1+\nu ^2-(1+\nu )^2zz^{\prime }+\nu (\nu
+1)(1-zz^{\prime })(z\frac \partial {\partial z}+z^{\prime }\frac \partial
{\partial z^{\prime }}) \\
&&-\nu ^2(1-zz^{\prime })^2\frac{\partial ^2}{\partial z\partial z^{\prime }}
]~\Phi _{M-1}(\gamma _1,\gamma _2)
\end{eqnarray*}

Using this equation we obtain for $\Phi _M(\gamma )$
\begin{eqnarray}
\Phi _M(\gamma ) &=&[1+\nu ^2-(1+\nu )^2\sin ^2\frac \gamma 2+\nu (\nu
+1)\sin \gamma \frac d{d\gamma }  \nonumber \\
&&-\nu ^2\cos ^2\frac \gamma 2(\frac{d^2}{d\gamma ^2}+\frac 1{\sin \gamma
}\frac d{d\gamma })]\Phi _{M-1}(\gamma )  \label{13}
\end{eqnarray}

The solution of Eq.(13) is
\begin{equation}
\Phi _M(y)=L^M(y)\Phi _0(y),  \label{14}
\end{equation}
where
\begin{eqnarray}
L(y) = \frac{(1-\nu )^2}2+\frac{(1+\nu )^2}2y-\nu (\nu +1)(1-y^2)\frac d{dy}
-\frac{\nu ^2}2(1+y)^2\frac d{dy}(1-y)\frac d{dy},  \label{15}
\end{eqnarray}
$y=\cos \gamma $ and $\Phi _0(y)=1.$

According to Eq.(15) $\Phi _M(y)$ is a polynomial in $y$ of order $M$. It
turns out that further calculations will be simplified if $\Phi _M(y)$ is
expanded over Legendre polynomials $P_n(y)$ :
\begin{equation}
\Phi _M(y)=\sum_{n=0}^Mc_n(M)P_n(y)  \label{16}
\end{equation}

The coefficients $c_n(l)$ are defined by the recurrent equation
\begin{eqnarray}
c_n(l+1) = \frac n{2n-1}\frac{(\nu n+1)^2}2c_{n-1}(l)+\frac{\nu
^2(n^2+n)+(\nu -1)^2}2c_n(l)
+\frac{n+1}{2n+3}\frac{(\nu n+\nu -1)^2}2c_{n+1}(l)  \label{17}
\end{eqnarray}
with initial condition $c_0(0)=1.$ Besides, $c_n(l)=0$ at $n>l.$

The norm $G_M$ is given by
\begin{equation}
G_M=\frac 12\int_{-1}^1\Phi _M(y)dy=c_0(M)  \label{18}
\end{equation}

\subsection{Spin correlations.}

The spin correlation functions can be found in the same way as $G_M.$ It is
convenient to express the scalar product ${\bf S}_i\cdot {\bf S}_j$ by the
permutation operator $p_{ij}=2{\bf S}_i\cdot {\bf S}_j+\frac 12$. Then the
non-normalized expectation value $Q(i,j)=$ $\langle \Psi _0(M)({\bf
S}_i\cdot{\bf S}_j-\frac 14)\Psi _0(M)\rangle $ is
\begin{eqnarray}
Q(i,j) =-\frac 14\langle \Psi _0(M)(p_{i,j}-1)^2\Psi _0(M)\rangle
= -\frac 14\langle \Psi _M(p_{i,j}-1)P_0(p_{i,j}-1)\Psi _M\rangle
\label{19}
\end{eqnarray}

The equations for $Q(i,j)$ can be obtained as in the derivation of
Eqs.(14)-(15). They are somewhat different for even and odd $i$. For
example, the equations for $Q(1,n)$ have forms
\begin{equation}
Q(1,2l)=-\frac{(\nu -1)^2}4\int_{-1}^1\frac{1+y}2(L^{l-2}L_1L^{M-l}~1)dy,
\label{20}
\end{equation}
\begin{equation}
Q(1,2l-1)=-\frac{\nu ^2(\nu -1)^2}4\int_{-1}^1\frac{1+y}
2(L^{l-2}L_2L^{M-l}~1)dy,  \label{21}
\end{equation}
where
\[
L_1=1-2\nu (1-y)\frac d{dy}-\nu ^2(1+y)\frac d{dy}(1-y)\frac d{dy} ,
\]
\[
L_2=1-2(1-y)\frac d{dy}-(1+y)\frac d{dy}(1-y)\frac d{dy}
\]

It is clear, that
\[
Q(2,n+2)=Q(1,N-n+1)
\]

Making use consequent integration of Eqs.(20) and (21), we obtain
\begin{eqnarray*}
\int_{-1}^1\frac{1+y}2(L^{l-2}L_{1(2)}L^{M-l}1)dy = \int_{-1}^1\frac{1+y}2(
\widetilde{L}^{l-2}1)(L_{1(2)}L^{M-l}1)dy
=\int_{-1}^1(\widetilde{L}_{1(2)}\widetilde{L}^{l-2}1)(L^{M-l}1)dy,
\end{eqnarray*}
where
\[
\widetilde{L}=\widetilde{L}_1=\frac 12+\frac{(2\nu -1)^2}2y-\nu (2\nu
-1)(1-y^2)\frac d{dy}-\frac{\nu ^2}2(1+y)^2\frac d{dy}(1-y)\frac d{dy} ,
\]
\[
\widetilde{L}_2=\frac{1+y}2-(1-y^2)\frac d{dy}-\frac{(1+y)^2}2\frac
d{dy}(1-y)\frac d{dy},
\]

So, Eqs.(20) and (21) can be rewritten as
\begin{equation}
Q(1,2l)=-\frac{(\nu -1)^2}4\int_{-1}^1\widetilde{\Phi }_{l-1}(y)~\Phi
_{M-l}(y)dy  , \label{22}
\end{equation}
\begin{equation}
Q(1,2l-1)=-\frac{\nu ^2(\nu -1)^2}4\int_{-1}^1(\widetilde{L}_2\widetilde{
\Phi }_{l-2}(y))~\Phi _{M-l}(y)dy , \label{23}
\end{equation}
where
\begin{equation}
\widetilde{\Phi }_l(y)=\widetilde{L}^l(y)~1  \label{24}
\end{equation}

The function $\widetilde{\Phi }_l(y)$ can be expanded over Legendre
polynomials similarly to $\Phi _l(y)$
\begin{equation}
\widetilde{\Phi }_l(y)=\sum_{n=0}^la_n(l)P_n(y) , \label{25}
\end{equation}
and coefficients $a_n(l)$ satisfy the equations
\begin{eqnarray}
a_n(l+1) = \frac n{2n-1}\frac{(\nu n+\nu -1)^2}2a_{n-1}(l)+\frac{\nu
^2(n^2+n)+1}2a_n(l)
+\frac{n+1}{2n+3}\frac{(\nu n+1)^2}2a_{n+1}(l)  \label{26}
\end{eqnarray}
with initial condition $a_0(0)=1$ and $a_n(l)=0$ at $n>l.$

Using Eqs.(18), (22) and (23) we can express the correlation function $
K(1,n)=\langle {\bf S}_1\cdot {\bf S}_n\rangle /G_M$ in the forms
\begin{equation}
K(1,2l)=\frac 14-\frac{(\nu -1)^2}{2c_0(M)}\sum_{n=0}^{l-1}\frac{
a_n(l-1)c_n(M-l)}{2n+1} , \label{27}
\end{equation}
\begin{equation}
K(1,2l-1)=\frac 14-\frac{\nu ^2(\nu -1)^2}{2c_0(M)}\sum_{n=0}^{l-1}\frac{
b_n(l-1)c_n(M-l)}{2n+1} , \label{28}
\end{equation}
where coefficients $b_n(l-1)$ are defined by $a_n(l-2)$ as follows:
\begin{equation}
b_n(l-1)=\frac 12\frac{n^3}{2n-1}a_{n-1}(l-2)+\frac{n^2+n+1}2a_n(l-2)+\frac
12\frac{(n+1)^3}{2n+3}a_{n+1}(l-2)  \label{29}
\end{equation}

Therefore, the calculation of the spin correlation function reduces to the
solution of the recurrent equations (17) and (26) which were used for
numerical calculations of the spin correlation function for finite systems.

At large $M$ the solutions of the recurrent equations (17) and (26) have
scaling forms:
\begin{equation}
c_n(M)=2M(M!)^2\nu ^{2M}s\exp \left(Mf_0(s)+\frac{2-\nu }\nu \ln M+f_1(s)\right),
\label{30}
\end{equation}
\begin{equation}
a_n(M)=2M(M!)^2\nu ^{2M}s\exp \left(Mf_0(s)-\frac{2-\nu }\nu \ln M+g_1(s)\right),
\label{31}
\end{equation}
where the parameter
\[
s=\frac{2n+1}{2M}
\]
can be considered as a continuous variable.

We note that Eqs.(30) and (31) are not valid for special values of $\nu $, $
\nu =\frac 1{m+1}$ ($m=0,1,2\ldots $). For these $\nu $ the last term in (17)
vanishes when $n=m$ and Eq.(17) reduces to ($m+1$) equations for $c_n(l)$
with $n\leq m$.

The recurrent equations (17) and (26) at $M>>1$ and $\nu \neq \frac 1{m+1}$
reduce to differential ones for $f_0(s)$, $f_1(s)$ and $g_1(s).$ For
example, $f_0(s)$ satisfies the equation
\[
\exp \left( f_0(s)-s\frac{df_0(s)}{ds} \right)
=s^2\cosh ^2\left(\frac 12\frac{df_0(s)}{ds}\right)
\]
with initial condition $f_0(1)=-\ln 4$.

Its implicit solution is
\begin{equation}
f_0(s)=-2\ln \xi +2\frac{\sin \xi }\xi \ln \tan \frac \xi 2,  \label{32}
\end{equation}
where
\[
s=\frac{\sin \xi }\xi
\]

As it follows from Eq.(32), $a_n(M)$ and $c_n(M)$ as a functions of $n$ have
a sharp maximum at $\frac nM=\frac 2\pi .$

The functions $f_1(s)$ and $g_1(s)$ are given by
\begin{equation}
f_1(s)=-\frac{2+\nu }\nu \ln \xi +\frac{4-\nu }{2\nu }\ln \sin \xi -\frac
12\ln (\frac{\sin \xi }{\xi ^2}-\frac{\cos \xi }\xi )+C\frac{\nu -1}{\nu ^2},
\label{33}
\end{equation}
\begin{equation}
g_1(s)=\frac{2-3\nu }\nu \ln \xi +\frac{3\nu -4}{2\nu }\ln \sin \xi -\frac
12\ln (\frac{\sin \xi }{\xi ^2}-\frac{\cos \xi }\xi )+C\frac{\nu -1}{\nu ^2},
\label{34}
\end{equation}
where constant $C\simeq 0.969$.

To obtain $K(1,2l)$ at $N\rightarrow \infty $ we substitute Eqs.(30)-(34)
into Eq.(27) and replace the sum over $n$ by the integral over $s.$ This
integral is calculated by the method of steepest descent. The saddle point
is $s_0=\frac 1\pi \sin (\frac{2\pi l}N)$ and the integrand does not depend
on parameter $\nu $. The final result for the spin correlation function $
K(1,2l)$ at $N\rightarrow \infty $ is also independent on $\nu $:
\begin{equation}
K(1,2l)=\frac 14\cos \left( \frac{2\pi (2l-1)}N \right)+O\left(\frac 1N\right)  \label{35}
\end{equation}

For the particular case $\nu =2$ Eq.(35) has been obtained earlier in Ref.18.

The corresponding calculations for $K(1,2l-1)$ to within terms $\sim \frac
1N $ lead to the same expression (35). But taking into account terms $\sim
\frac 1N$ we find that the difference $K(1,2l-1)-K(1,2l)$ is
\[
K(1,2l-1)-K(1,2l)=\frac \pi N\frac{\nu -1}\nu \sin\left(\frac{4\pi l}N\right)
+O\left(\frac 1{N^2}\right)
\]

The latter equation means that the double-spiral structure exists. The pitch
angle of each spiral is $\frac{4\pi }N$ and there is a small shift angle $
\triangle \varphi =\frac{2\pi }N\frac{2-\nu }\nu $ between them:
\begin{equation}
K(1,2l)=\frac 14\cos \left(\frac{2\pi (2l-1)}N-\triangle \varphi \right) , \label{36}
\end{equation}
\begin{equation}
K(1,2l+1)=\frac 14\cos\left( \frac{4\pi l}N \right) \label{37}
\end{equation}

This shift angle reflects the fact that the unit cell contains two sites
unless $\nu =2$.

Eqs.(36) and (37) show that the long range spiral order exists in the
singlet ground state of the Hamiltonian (2) in the thermodynamic limit and
the double-spiral state is formed.

It is interesting to note that correlators (36) and (37) coincide with those
obtained by using the simple `quasi-classical' trial wave function in a
form
\[
\psi _{cl}=\exp \left(\sum_{n=1}^N\varsigma _ns_n^{+}\right)
\left| 1,2,\ldots N\right\rangle,
\]
where
\[
\varsigma _n=\exp \left(\frac{2\pi n}N+\frac{(-1)^n\pi }N\frac{\nu -2}\nu \right)
\]

Thus, the quantum ground state of the large-$N$ limit resembles the
classical one though for small size systems the quantum fluctuations are
essential.

The formation of the spirals having the period which is equal to the system
size reflects the tendency to the creation of the incommensurate spiral
state at the antiferromagnetic region when $\delta >0$. The behavior of the
system in the vicinity of the transition point has been studied by us$^7$
for the model (1) with $J_{12}=J_{23}=-1$, $J_{13}=\frac 14+\delta $. For $
\delta \ll 1$ the period of the spiral is finite and is proportional to $
\delta ^{-\frac 12}$. The transition from the ferromagnetic to the singlet
state is a phase transition of the second order with respect to $\delta $.

\subsection{Special cases of the model.}

There are the special points, $\nu =\frac 1m$ ($m=1,2,\ldots $) at which
Eqs.(36) and (37) are not valid.

At $\nu =\frac 12$ the function (4) reduces to the product of singlets
\begin{equation}
\Psi _0(M)=[2,3][4,5]\ldots [N,1] , \label{38}
\end{equation}
and $K(2,3)=K(4,5)=\ldots =K(N,1)=-\frac 34$. Other spin correlators are zero.

Analysis of Eqs.(17), (26) and (27) shows that the ground state correlations
of the model (2) with $\nu =\frac 1m$ ($m\geq 3$) have antiferromagnetic
character with an exponentially decay:
\[
K(1,n+1)\sim (-1)^n\exp (-\frac n{r_c}),
\]
where the correlation length is
\begin{equation}
r_c=-2\ln ^{-1}\left( 1-\frac 2{m(m-1)}\right)  \label{39}
\end{equation}

The crossover between the spiral state at $\frac 1m<\nu <\frac 1{m-1}$ and
the antiferromagnetic state at $\nu =\frac 1m$ occurs in the exponentially
small (at $N\gg 1$) vicinity of these spacial points. At $m\gg 1$
\begin{equation}
r_c=m^2  , \label{40}
\end{equation}
and the correlation length diverges when $\nu $ trends to zero along the
special points and there is the Neel ordering in this limit.

At $\nu =1$, when $J_{12}=-1$, $J_{23}=J_{13}=0$, the system is divided into
the ferromagnetic pairs $1-2,3-4,\ldots $.

Using the second order perturbation theory with respect to $(\nu -1)$, we
reduce the model (2) at $\nu \rightarrow 1$ to the effective spin-$1$
Hamiltonian
\begin{eqnarray}
H_{eff} &=&\frac{(\nu -1)^2}8\sum_{n=1}^M\left\{ -({\bf L}_n\cdot{\bf 
L}_{n+1}-1)+{\bf L}_n\cdot{\bf L}_{n+2}\right.  \nonumber \\
&&\left. -\frac 12({\bf L}_n\cdot{\bf L}_{n+1})({\bf L}_{n+1}\cdot{\bf L}
_{n+2})-\frac 12({\bf L}_{n+1}\cdot{\bf L}_{n+2})({\bf L}_n\cdot{\bf
L}_{n+1})\right\}, 
\label{41}
\end{eqnarray}
with $M$ spins $L=1$.

The exact singlet ground state wave function of the Hamiltonian (41) can be
obtained from the Eqs.(4) and (5). It has a form
\begin{equation}
\Psi
_0(M)=P_0~L_1^{+}(L_2^{+}+L_3^{+}+...+L_M^{+})\ldots
(L_{M-1}^{+}+L_M^{+})L_M^{+}~|-1,\ldots ,-1\rangle ,
\label{42}
\end{equation}
where $L_i^{+}$ are raising operators of spin $1$.

The ground state correlation function of (41) $K(1,n+1)=\langle {\bf L}_1\cdot
{\bf L}_{n+1}\rangle /G_M$ is found from Eqs.(36) and (37). It is
\[
K(1,n+1)=\cos \left( \frac{2\pi n}M \right)
\]

Finally, we note that it is possible to calculate higher terms of the
perturbation theory and to obtain an effective spin-$1$ Hamiltonians which
are proportional to the third, fourth and higher power of the small
parameter $(\nu -1)$. All of them have zero ground state energy as well as
(41).

\newpage
\section{Frustrated spin chain with an antiferromagnetic ground state.}

\subsection{The model and its exact ground state.}

In the preceding section the spin model at the ferromagnet-antiferromagnet
transition point has been studied. The exact singlet ground state wave
function at this point is given by Eq.(4). In this section the function (4)
will be generalized to give the exact ground state wave function of new
special one-dimensional frustrated spin-$\frac 12$ model. This model has the
unique singlet ground state (for the cyclic chain) with an exponentially
decay of correlations and there is a gap to the excitations.

Let us consider the wave function which depends on two parameters and has
the form (4)
\begin{equation}
\Psi _0(M)=P_0\Psi _M,  \label{43}
\end{equation}
where
\begin{eqnarray}
\Psi _M &=&(s_1^{+}+\nu _1s_2^{+}+\nu _2s_3^{+}\ldots +\nu
_2s_N^{+})(s_3^{+}+\nu _1s_4^{+}+\nu _2s_5^{+}\ldots +\nu _2s_N^{+})\ldots 
\nonumber \\
&&\times (s_{N-1}^{+}+\nu _1s_N^{+})~\left| 1,2,\ldots N\right\rangle  \label{44}
\end{eqnarray}

We will construct the Hamiltonian for which $\Psi _0(M)$ is the exact ground
state wave function as the sum of the local four-sites Hamiltonians

\begin{equation}
H=H_{1,2,3,4}+H_{3,4,5,6}+\ldots +H_{N-3,N-2,N-1,N}+H_{N-1,N,1,2}  \label{45}
\end{equation}

The Hamiltonians $H_{i,i+1,i+2,i+3}$ are chosen in the form:

\begin{eqnarray}
H_{i,i+1,i+2,i+3} &=&J_{12}\left[ ({\bf S}_i\cdot {\bf S}_{i+1}-\frac 14)+({\bf S}
_{i+2}\cdot {\bf S}_{i+3}-\frac 14)\right]  \nonumber \\
&&+2J_{13}\left[ ({\bf S}_i\cdot {\bf S}_{i+2}-\frac 14)+({\bf S}_{i+1}\cdot {\bf S}
_{i+3}-\frac 14)\right]  \nonumber \\
&&+2J_{23}({\bf S}_{i+1}\cdot {\bf S}_{i+2}-\frac 14)+2J_{14}({\bf S}_i\cdot {\bf S}
_{i+3}-\frac 14)  \label{46}
\end{eqnarray}

Thus, the model has nearest ($J_{12}$ and $J_{23}$), next-nearest ($J_{13}$)
and next-next-nearest ($J_{14}$) neighbor interactions. In fact, this model
is equivalent to the spin-$\frac 12$ ladder with different interactions as
it is shown in Fig.2.

\begin{figure}[t]
\unitlength1cm
\begin{picture}(11,4)
\centerline{\psfig{file=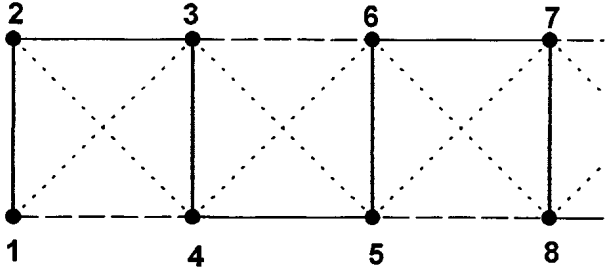,angle=0,width=9cm}}
\end{picture}
\caption[]{Spin ladder representation of the model (45). Different lines
correspond to different exchange interactions.}
\end{figure}

We demand that $\Psi _0(M)$ is the eigenfunction of each local Hamiltonian
with the eigenvalue $\varepsilon <0$, i.e.
\begin{equation}
H_{i,i+1,i+2,i+3}\Psi _0(M)=\varepsilon ~\Psi _0(M)  \label{47}
\end{equation}

The exchange integrals $J_{ij}$ and the energy $\varepsilon $ is defined by
the Schr\"odinger equation:
\begin{equation}
H_{1,2,3,4}\Psi _M=\varepsilon ~\Psi _M  \label{48}
\end{equation}

Let us represent the function $\Psi _M$ in the form
\begin{equation}
\Psi _M=\left[ \widehat{q}_1+\nu _2\widehat{q}_2S^{+}(N-4)+\nu
_2^2S^{+2}(N-4)\right] \left| 1,2,3,4\right\rangle \Psi _{M-2},  \label{49}
\end{equation}
where
\[
\widehat{q}_1=s_1^{+}s_3^{+}+\nu _1(s_1^{+}s_4^{+}+s_2^{+}s_3^{+})+\nu
_1^2s_2^{+}s_4^{+}+\nu _2(1+\nu _1)s_3^{+}s_4^{+} ,
\]
\[
\widehat{q}_2=s_1^{+}+\nu _1s_2^{+}+(1+\nu _2)s_3^{+}+(\nu _1+\nu _2)s_4^{+}
\]

The equation
\begin{equation}
H_{1,2,3,4}\widehat{q}_1\left| 1,2,3,4\right\rangle =\varepsilon ~\widehat{q}
_1\left| 1,2,3,4\right\rangle  \label{50}
\end{equation}
reduces to five equations for four exchange integrals and the energy $
\varepsilon $. The necessary condition of the existence of a solution with $
\varepsilon <0$ is
\begin{equation}
1+\nu _1+\nu _2=0  \label{51}
\end{equation}

So, under this condition there is only one parameter of the Hamiltonian
(45). It is convenient to take the value $\mu =\frac{2+\nu _1}{2\nu _1+1}$
as a system parameter. Then, the solution of Eq.(50) at $\nu _1=\frac{2-\mu
}{2\mu -1}$ and $\nu _2=\frac{\mu +1}{1-2\mu }$ yields the following
expressions for $J_{ij}$ and $\varepsilon $:
\begin{equation}
\begin{array}{c}
J_{12}=\frac{(1-2\mu )(2-\mu )}9,\qquad J_{23}=-\frac 13\frac{(2-\mu )^2(\mu
-1)\mu }{(\mu +1)^2} \\
\\
J_{13}=\frac{(2\mu -1)(2-\mu )(\mu -1)^2}{6(\mu +1)^2},\qquad J_{14}=\frac{
(2\mu -1)^2(\mu -1)}{3(\mu +1)^2} \\
\\
\varepsilon =-\frac{3\mu (\mu -1)^2}{(\mu +1)^2}
\end{array}
\label{52}
\end{equation}

It turns out that the equation
\[
H_{1,2,3,4}\widehat{q}_2\left| 1,2,3,4\right\rangle =\varepsilon ~\widehat{q}
_2\left| 1,2,3,4\right\rangle
\]
with $J_{ij}$ and $\varepsilon $ given by Eq.(52) is satisfied automatically.

As it will be proven in Appendix the function $\Psi _M$ with $\nu _1=-\nu
_2-1$ contains the singlet and the triplet components only, i.e.
\begin{equation}
S^{+2}(N)\Psi _M=0  \label{53}
\end{equation}

Therefore, the last term in Eq.(49) vanishes and $\Psi _M$ is the
eigenfunction of $H_{1,2,3,4}$ with the eigenvalue $\varepsilon $.

Generally, the Hamiltonian $H_{1,2,3,4}$ has following eigenstates: one
quintet, three triplets and two singlets. Two of them (one singlet and one
triplet) have the energy $\varepsilon $ while other four states have higher
energies at $\mu >0$ ($\mu \neq 1$). At $\mu <0$ the ground state of $
H_{1,2,3,4}$ is a quintet with zero energy. In Appendix we will show that
the wave function $\Psi _M$ is the eigenfunction with the eigenvalue $
\varepsilon $ of each local Hamiltonian $H_{i,i+1,i+2,i+3}$ excluding that
for $i=N-1$. Therefore, the ground state energy of the open chain at $\mu >0$
is $(\frac N2-1)\varepsilon $ and it coincides with the exact lower bound of
the energy similarly to the model of Section 2. However, in the contrast
with the latter the present model is four-fold degenerate for the open chain.

As for the Hamiltonian $H_{N-1,N,1,2}$ the function $\Psi _M$ is not its
eigenfunction, but
\[
H_{N-1,N,1,2}\Psi _0(M)=\varepsilon ~\Psi _0(M)
\]
(see Appendix) and
\[
H\Psi _0(M)=\frac N2\varepsilon ~\Psi _0(M)
\]

As it follows from Eq.(52) the spectrum of $H(\mu )$ coincides with the
spectrum of $\widetilde{H}(\mu )$ which is connected with $H(\mu )$ by
transformation
\[
\widetilde{H}(\mu )=\mu ^2H(\frac 1\mu )
\]

This transformation permutes the factors at the third and the last terms in
Eq.(46). Therefore, it is sufficient to consider the Hamiltonian $H(\mu )$
in the region $-1\leq \mu \leq 1$.

The ground state of $H$ is ferromagnetic at $\mu <0$. When $0<\mu <1$ the
ground state of the cyclic chain is the non-degenerate singlet. The point $
\mu =0$ is the ferromagnet-antiferromagnet transition point. At this point
the present model coincides with the model (2) at the special point $\nu
=\frac 13$. As it follows from Eq.(52) this transition is the phase
transition of the first order with respect to $\mu $.

At $\mu =\frac 12$ the only non-zero exchange integral is $J_{23}$ and the
ground state consists of non-interacting singlet pairs $2-3,4-5,\ldots ,1-N$.
When $\mu \rightarrow 1$ the first term in Eq.(46) dominates and the system
is divided into weekly interacting ferromagnetic pairs $1-2,3-4,\ldots $ . Using
the second order perturbation theory with respect to the small parameter $
(\mu -1)$ we reduce the Hamiltonian $H$ (Eq.45) to the effective spin-$1$
model given by
\[
H=\frac{(\mu -1)^2}{16}H_{eff},
\]
where
\begin{equation}
H_{eff}=\sum_{n=1}^M\left\{ 5{\bf L}_n\cdot {\bf L}_{n+1}+{\bf L}_n\cdot 
{\bf L}_{n+2}+\frac 12({\bf L}_n\cdot {\bf L}_{n+1}-{\bf L}_{n+1}\cdot {\bf L}
_{n+2})^2-6\right\} , \label{54}
\end{equation}
and ${\bf L}_n$ is spin-$1$ operator.

The ground state wave function of (54) can be obtained from Eq.(43) at $\mu
=1$. It reads
\begin{eqnarray}
\Psi _0(M) =P_0~({\bf L}_1^{+}-2{\bf L}_2^{+}-\ldots -2{\bf L}_M^{+})({\bf L}
_2^{+}-2{\bf L}_3^{+}-\ldots -2{\bf L}_M^{+})\ldots \times 
({\bf L}_{M-1}^{+}-2{\bf L}_M^{+}){\bf L}_M^{+}~\left|
-1,-1,\ldots -1\right\rangle  \label{55}
\end{eqnarray}

It is remarkable that the function (55) coincides with the ground state wave
function of the AKLT model. Therefore, the ground state physics of the model
given by Eq.(54) and AKLT one is the same, though the Hamiltonians of these
two models are different.

\subsection{Spin correlations in the ground state.}

First, we calculate the norm of the ground state wave function. It is
convenient to express the function $\Psi _M$ in terms of the parameter $\mu $
and to introduce a new function $\widetilde{\Psi }_M$
\[
\Psi _M=(2\mu -1)^{-M}~\widetilde{\Psi }_M,
\]
where
\begin{eqnarray*}
\widetilde{\Psi }_M &=&\left[ (2\mu -1)s_1^{+}+(2-\mu )s_2^{+}-(\mu
+1)S^{+}(N-2)\right]
\left[ (2\mu -1)s_3^{+}+(2-\mu )s_4^{+}-(\mu +1)S^{+}(N-4)\right] \\
&&\ldots \left[ (2\mu -1)s_{N-1}^{+}+(2-\mu )s_N^{+}\right] \left|
1,2,\ldots N\right\rangle
\end{eqnarray*}

According to Eq. (18) the norm of $\widetilde{\Psi }_0(M)=P_0\widetilde{\Psi
}_M$ is
\[
G_M=\left\langle \widetilde{\Psi }_0(M)\widetilde{\Psi }_0(M)\right\rangle
=\frac 12\int_{-1}^1\Phi _M(y)dy
\]

For the present model the function $\Phi _l(y)$ is defined by the equation
\begin{equation}
\Phi _l(y)=L^l(y)~1 , \label{56}
\end{equation}
where
\[
L(y)=\frac{9(1-\mu )^2}2+\frac{(1+\mu )^2}2\left[ y+2(1-y^2)\frac
d{dy}-(1+y)^2\frac d{dy}(1-y)\frac d{dy}\right]
\]

The solution of Eq.(56) is
\[
\Phi _l(y)=\frac{\omega _1^l+3\omega _2^l}4+\frac{\omega _1^l-\omega _2^l}4y,
\]
where
\[
\omega _1=6(\mu ^2-\mu +1),\qquad \omega _2=2(\mu -2)(2\mu -1)
\]

This form of $\Phi _l(y)$ reflects the fact that the function $\widetilde{
\Psi }_M$ contains the singlet and the triplet components only. Thus, $G_M$ is
\begin{equation}
G_M=\frac{\omega _1^M+3\omega _2^M}4  \label{57}
\end{equation}

As $\left| \frac{\omega _2}{\omega _1}\right| <1$ then $G_M=\frac 14\omega
_1^M$ at $M\rightarrow \infty $.

The spin correlation functions can be found in the similar way as in the
Section 2. In analogy with the Eqs.(19)-(20), we obtain
\begin{equation}
Q(N-2l-1,N)=-\frac 18\int_{-1}^1(L^{M-l-1}\Phi _{l+1}^{\prime }(y))dy,
\label{58}
\end{equation}
where
\[
\Phi _{l+1}^{\prime }(y)=\langle \widetilde{\Psi }
_{l+1}(p_{N-2l-1,N}-1)P_0(p_{N-2l-1,N}-1)\widetilde{\Psi }_{l+1}\rangle
\]

Carrying out the necessary calculations, we find
\begin{eqnarray*}
\Phi _{l+1}^{\prime }(y) =4(\mu +1)^2(2-\mu )^2\omega _1^{l-1}+\omega
_2^{l+1}+(2\mu -1)^2 
\left[ (2\mu ^2-2\mu +5)\omega _1^{l-1}+(2\mu ^2+10\mu -1)\omega
_2^{l-1}\right] (1+y)
\end{eqnarray*}

Substituting $\Phi _{l+1}^{\prime }(y)$ into Eq.(58) we obtain for the spin
correlation function $K(1,2l+2)=K(N-2l-1,N)$
\begin{equation}
K(1,2)=\frac 14-\frac{9(\mu -1)^2}{2\omega _1}\frac{1+3\left( \frac{\omega _2
}{\omega _1}\right) ^{M-1}}{1+3\left( \frac{\omega _2}{\omega _1}\right) ^M},
\label{59}
\end{equation}
\begin{equation}
K(1,2l+2)=-\frac{3(\mu +1)^2}{\omega _1\omega _2}\frac{(2\mu -1)^2\left(
\frac{\omega _2}{\omega _1}\right) ^l+(\mu -2)^2\left( \frac{\omega _2}{
\omega _1}\right) ^{M-l}}{1+3\left( \frac{\omega _2}{\omega _1}\right) ^M},
\label{60}
\end{equation}
where $l=1,2,\ldots M-1$.

The similar calculations for $K(1,2l+1)$ result in
\begin{equation}
K(1,2l+1)=\frac{3(\mu +1)^2}{2\omega _1}\frac{\left( \frac{\omega _2}{\omega
_1}\right) ^l+\left( \frac{\omega _2}{\omega _1}\right) ^{M-l}}{1+3\left(
\frac{\omega _2}{\omega _1}\right) ^M}  \label{61}
\end{equation}

The correlators $K(2,n)$ have been obtained by using the symmetry of the
system (see Fig.2). For example,
\[
K(2,3)=K(1,N),\quad etc.
\]

In the thermodynamic limit $M\rightarrow \infty $ and $l\ll M$ Eqs.(59)-(61)
reduce to
\begin{equation}
K(1,2)=-\frac 34\frac{\omega _2}{\omega _1} , \label{62}
\end{equation}
\begin{equation}
K(1,2l+2)=-\frac{3(\mu +1)^2(2\mu -1)^2}{\omega _1\omega _2}\left( \frac{
\omega _2}{\omega _1}\right) ^l  \label{63}
\end{equation}
\begin{equation}
K(1,2l+1)=\frac{3(\mu +1)^2}{2\omega _1}\left( \frac{\omega _2}{\omega _1}
\right) ^l  \label{64}
\end{equation}

The correlators have the exponential decay and the correlation length $r_c$
is
\begin{equation}
r_c(\mu )=2\ln ^{-1}\left| \frac{\omega _1(\mu )}{\omega _2(\mu )}\right|
\label{65}
\end{equation}

It follows from Eq.(65) that the ground state has ultrashort-range
correlations. For example, $r_c(0)=2\log ^{-1}\left( \frac 32\right) $ (this
value coincides with $r_c$ given by Eq.(39) with $m=3$) and $r_c(1)=2\log
^{-1}3$. In the latter case $r_c$ coincides with correlation length of the
AKLT\ model. At $\mu =\frac 12$ the only non-zero correlators are $
K(2,3)=K(4,5)=\ldots =K(N,1)=-\frac 34$ in accordance with the dimer character
of the ground state.

The value $\omega _2(\mu )$ changes the sign at $\mu =\frac 12$ and as it
follows from Eqs.(62), (63) and (64) the correlators show the
antiferromagnetic structure of the ground state at $0\leq \mu \leq \frac 12$
while at $\frac 12\leq \mu \leq 1$ there are ferromagnetic correlations
inside pairs $(1,2), (3,4),\ldots $ and the antiferromagnetic correlations
between the pairs.

\subsection{Energy gap.}

The Hamiltonian $H$ of the cyclic chain has a singlet-triplet gap $\Delta $
for finite $N$. It is evident that for $\mu =\frac 12$ the gap exists for $
N\rightarrow \infty $ and $\Delta (\frac 12)=\frac 16$. The existence of the
finite gap at the thermodynamic limit in the range $0<\mu <1$ follows from
the continuity of the function $\Delta (\mu )$. It is also clear that $
\Delta (\mu )$ at $N\rightarrow \infty $ vanishes at the boundary points $
\mu =0$ and $\mu =1$ when the ground state is degenerate and there are
low-lying spin wave excitations.

Unfortunately, the method of the exact calculation of $\Delta (\mu )$ in the
thermodynamic limit is unknown. For $\mu \simeq \frac 12$ the gap can be
found by using the perturbation theory in $(\mu -\frac 12)$. In this case $
\Delta (\mu )$ is
\begin{equation}
\begin{array}{c}
\Delta (\mu )=\frac 16+O((\mu -\frac 12)^2),\quad \mu \leq \frac 12 \\
\\
\Delta (\mu )=\frac 16-\frac 89(\mu -\frac 12)+O((\mu -\frac 12)^2),\quad
\mu \geq \frac 12
\end{array}
\label{66}
\end{equation}

Eq.(66) shows that $\Delta (\mu )$ has a cusp at $\mu =\frac 12$. For the
approximate calculation $\Delta (\mu )$ we use the trial function of the
triplet state in the form
\begin{equation}
\Psi _t = \sum_{n=1}^Nc_ns_n^{+}\Psi _0(M) , \label{67}
\end{equation}
where
\[
c_{2l-1}=ae^{ikl},\quad c_{2l}=be^{ikl},\quad k=\frac{4\pi }Nt,\quad
t=1,\ldots M
\]

This trial function gives $\Delta (\mu )$ which is
\begin{equation}
\Delta (\mu )=\frac { 2\sum_{n\neq m} c_n^{*} c_m K(n,m) J_{n,m}
- 2 \sum_{n\neq m} |c_n^2| K(n,m) J_{n,m} }
{ \frac 34 \sum_n |c_n^2| + \sum_{n\neq m} c_n^{*} c_m K(n,m)}  \label{68}
\end{equation}

Function (68) has minima at $k=\frac{4\pi }N$ and $k=\pi $ for $0<\mu <\frac
12$ and $\frac 12<\mu <1$, respectively. Then $\Delta (\mu )$ at $
N\rightarrow \infty $ is
\begin{equation}
\begin{array}{c}
\Delta (\mu )=\frac 83 \frac{(\mu +1)^4(\mu -1)^2}{\omega _1(\omega
_1+\omega _2)},\qquad 0<\mu <\frac 12 \\
\\
\Delta (\mu )=\frac 23(1+\frac{\omega _2}{\omega _1})(\mu -1)^2,\qquad \frac
12<\mu <1
\end{array}
\label{69}
\end{equation}

The dependance of $\Delta (\mu )$ given by Eq.(69) is shown on Fig.3
together with the results of extrapolations of exact finite-chain
calculations. Both dependences agree very well for $\mu \geq \frac 12$. In
particular, Eq.(69) correctly reproduces (66) at $\mu \simeq \frac 12$.
However, $\Delta (\mu )$ given by Eq.(69) is not zero at $\mu =0$ while
numerical calculations fit the dependence $\Delta (\mu )\sim \sqrt{\mu }$ at
$\mu \rightarrow 0$.

\begin{figure}[t]
\unitlength1cm
\begin{picture}(11,8)
\centerline{\psfig{file=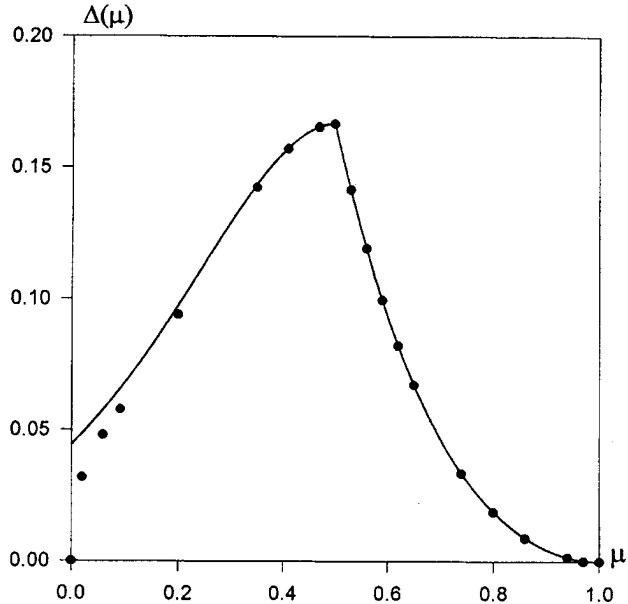,angle=0,width=9cm}}
\end{picture}
\caption[]{Singlet-triplet gap of the model (45) as a function of the
parameter $\mu$. The circles denote the results of the extrapolation of an
exact finite-chain calculations. The solid line represents the dependence
$\Delta (\mu )$ given by Eq.(69).}
\end{figure}

We note that the trial function of the type (67) gives the value $0.7407$
for the singlet-triplet gap in the AKLT\ model. This estimate is close to
the value $0.7143$ obtained by another approach in Ref.20.

The above consideration refers to the gap in the cyclic chain. The open
chain has four-fold degenerate ground state. Finite chain calculations show
that the spin of the lowest excitation is $S=2$ . However, there are also
two excited singlet and triplet states the energies of which are close to
that of $S=2$. The difference of these three eigenvalues decreases to zero
exponentially at $N\rightarrow \infty $. We expect that these states are
degenerate in the thermodynamic limit (at $\mu =\frac 12$ they are
degenerate for finite $N$). The gap in the open chain equals to the
difference between the energies of the degenerate ground state and the
lowest excited one. The extrapolation of the results of finite-chain
calculations to $N\rightarrow \infty $ gives the finite gap in the open
chains at $0<\mu <1$. Its value is very close to that for the cyclic chains.

\section{Summary}

We have studied a class of the one dimensional quantum spin-$\frac 12$
models with competing interactions. The exact ground state wave function of
these models is found in the special recurrent form. The Hamiltonians of
these models are the sums of local, non-commuting with each other
Hamiltonians; the ground state wave function of the total Hamiltonian is the
ground state solution for each of them. This means that this ground state
wave function is superstable$^{19}$ with respect to each local Hamiltonian.

One of the studied models describes the transition point from the
ferromagnetic to the spiral state when the energies of these two states are
equal to each other. It is interesting to compare the exact quantum singlet
state with the classical one. Both states are the states of a helical type
(excluding some special cases) and their spin correlation functions are
identical in the thermodynamic limit though quantum effects are essential
for finite chains. This fact is rather surprising for the one-dimensional
model with spin $\frac 12$.

Another exactly solvable Hamiltonian has the antiferromagnetic ground state.
This state is non-degenerate for the closed chains and is four-fold
degenerate for the open ones. The Hamiltonian depends on the one parameter $
\mu $ and there are two special values, $\mu =0$ and $\mu =1$ where the
singlet and the ferromagnetic states are degenerate. The value $\mu =0$ is
the ferromagnet-antiferromagnet transition point where the phase transition
of the first order with respect to $\mu $ occurs.

The ground state is characterized by the exponential decay of correlators
with a very short correlation length, and there is the gap in the excitation
spectrum at $0<\mu <1$. Thus, this model has all properties suggested by
Haldane$^{16}$ for the one-dimensional Heisenberg antiferromagnet with
integer spin. The first model for which all these properties have been
proved rigorously is the AKLT model. Our model is the one with spin $\frac
12 $. Affleck and Lieb$^{21}$ have shown for the translationally invariant
and the isotropic Heisenberg Hamiltonians that for half-integer spin chain
either the excitation spectrum is gapless or the ground state is degenerate.
The existence of the finite gap in our model does not contradict to
Affleck-Lieb theorem because this model is not translationally invariant. It
has two sites in the unit cell and is equivalent to the special ladder
model. Moreover, in the limit $\mu \rightarrow 1$ its ground state wave
function reduces to that for the AKLT model.

\section{Acknowledgments}

We are grateful to Profs.M.Ya.Ovchinnikova and V.N.Prigodin for helpful
discussions. This work was supported by ISTC under Grant No.015-94 and in
part by RFFR No.33727.

\section{Appendix}

We prove that square of the total raising operator annihilates $\Psi _M$ if
the condition (51) is satisfied. The recurrent equation for $S^{+2}(N)\Psi
_M $ is
\begin{equation}
\begin{array}{c}
S^{+2}(N)\Psi _M=2(1+\nu _1+\nu _2)s_1^{+}s_2^{+}|1,2\rangle ~S^{+}(N-2)\Psi
_{M-1}+ \\
\\
+\left[ (1+2\nu _2)s_1^{+}+(\nu _1+2\nu _2)s_2^{+}+\nu _2S^{+}(N-2)\right]
|1,2\rangle ~S^{+2}(N-2)\Psi _{M-1}
\end{array}
\tag{A1}
\end{equation}

The first term in Eq.(A1) vanishes under the condition (51) and, therefore,
\begin{eqnarray*}
&&S^{+2}(N)\Psi _M=\left[ (1+2\nu _2)s_1^{+}+(\nu _1+2\nu _2)s_2^{+}+\nu
_2S^{+}(N-2)\right] \\
&&\times \left[ (1+2\nu _2)s_3^{+}+(\nu _1+2\nu _2)s_4^{+}+\nu
_2S^{+}(N-4)\right]\ldots 
\times (s_{N-1}^{+}+s_N^{+})^2\left[ s_{N-1}^{+}+\nu _1s_N^{+}\right]
|1,2,\ldots N\rangle =0
\end{eqnarray*}

This equation means that the wave function $\Psi _M$ contains the singlet
and triplet components only.

Now we prove that $\widetilde{\Psi }_0(M)$ is the eigenfunction of each
local Hamiltonian in (45). Of course, consequentely the same will be true
for $\Psi _0(M)$.

The function $\widetilde{\Psi }_M$ satisfies the recurrent equation
\begin{equation}
\widetilde{\Psi }_M=\left( \widehat{u}_{12}\widetilde{\Psi }_{M-1}-(\mu
+1)S^{+}(N-2)\widetilde{\Psi }_{M-1}\right) |1,2\rangle ,  \tag{A2}
\end{equation}
where
\[
\widehat{u}_{12}=(2\mu -1)s_1^{+}+(2-\mu )s_2^{+}
\]

Let us consider functions $\varphi _M^{+}=S^{+}(N)\widetilde{\Psi }_M$, $
\varphi _M^{-}=S^{-}(N)\widetilde{\Psi }_M$ and $\chi _M=S^{-}(N)S^{+}(N)
\widetilde{\Psi }_M$. The recurrent equations for these functions are
obtained from (A2) using (A1). They are
\begin{equation}
\varphi _M^{+}=\left( (\mu +1)s_1^{+}s_2^{+}\widetilde{\Psi }_{M-1}+\widehat{
v}_{12}\varphi _{M-1}^{+}\right) |1,2\rangle , \nonumber
\end{equation}
\begin{equation}
\varphi _M^{-}=\left( (\mu +1)(\widetilde{\Psi }_{M-1}-\chi _{M-1})+\widehat{
u}_{12}\varphi _{M-1}^{-}\right) |1,2\rangle ,\tag{A3}
\end{equation}
\begin{equation}
\chi _M=\left[ (\widehat{u}_{12}-\widehat{v}_{12})\widetilde{\Psi }_{M-1}+
\widehat{v}_{12}\chi _{M-1} + (\mu +1)(s_1^{+}s_2^{+}\varphi
_{M-1}^{-}-\varphi _{M-1}^{+})\right] |1,2\rangle \nonumber,
\end{equation}
where
\[
\widehat{v}_{12}=-(2-\mu )s_1^{+}-(2\mu -1)s_2^{+}
\]

Eqs.(A2)-(A3) can be written in a matrix form
\begin{equation}
R(M)=D_{12}R(M-1),  \tag{A4}
\end{equation}
where $R(M)$ and $D_{12}$ are $(2\times 2)$ matrices
\[
R(M)=\left(
\begin{array}{cc}
\widetilde{\Psi }_M-\chi _M & \varphi _M^{+} \\
-\varphi _M^{-} & \widetilde{\Psi }_M
\end{array}
\right) ,
\]
\[
D_{12}=\left(
\begin{array}{cc}
\widehat{v}_{12} & (\mu +1)s_1^{+}s_2^{+} \\
-(\mu +1) & \widehat{u}_{12}
\end{array}
\right) |1,2\rangle
\]

Therefore, $R(M)$ is
\begin{equation}
R(M)=D_{12}D_{34}\ldots  D_{N-1,N}  \tag{A5}
\end{equation}

As $\widetilde{\Psi }_M$ contains the singlet and triplet components only,
the projection of $\widetilde{\Psi }_M$ onto the singlet is
\begin{equation}
\widetilde{\Psi }_0(M)=P_0\widetilde{\Psi }_M=2\widetilde{\Psi }_M-\chi _M
\tag{A6}
\end{equation}

It follows from Eqs.(A5)-(A6) that
\begin{equation}
\widetilde{\Psi }_0(M)=Tr(D_{12}D_{34}\ldots D_{N-1,N})  \tag{A7}
\end{equation}

This form of $\widetilde{\Psi }_0(M)$ is similar to the matrix product wave
function of the AKLT model and its generalizations which has been found in
Ref.15. Each of four matrix elements of $R(M)$ is the eigenfunction of the
local Hamiltonian $H_{i,i+1,i+2,i+3}$ for $i=1,3,\ldots N-3$ because of matrix
elements of the product $D_{i,i+1}D_{i+1,i+2}$ are the eigenfunctions of
this Hamiltonian. Besides, it can be proved$^{15}$ that the four matrix
elements of $R(M)$ are the only ground states of (45) and, therefore, the
ground state of the open chain is four-fold degenerate.

It is easily to check$^{15}$ that the triplet wave functions $R_{12}(M)$ and
$R_{21}(M)$ are not eigenfunctions of $H_{N-1,N,1,2}$. On the other hand,
using cyclic permutations of matrices under the $trace$, we have
\[
H_{N-1,N,1,2}\widetilde{\Psi }_0(M)=\varepsilon ~\widetilde{\Psi }_0(M)
\]
and, therefore, the ground state of the cyclic chain is the non-degenerate
singlet.

\end{document}